\documentclass[11pt]{article}
\usepackage[left=2cm, right=2cm, top=2.5cm, bottom=2.5cm]{geometry}
\usepackage{float}
\usepackage{cite}
\usepackage[x11names,dvipsnames]{xcolor}
\usepackage{fancyhdr, amssymb, amsmath, graphicx, pgfplots, tikz}
\usepackage{isomath}
\usepackage{physics}
\usepackage{comment}

\usetikzlibrary{shadows}

\newcommand{\scolor}{blue!40!black}

\newcommand{\tcolor}{Orange!80!black}

\usepackage[labelfont={bf,sf, color=\scolor}, margin={1.5cm,0cm}]{caption}
\usepackage[colorlinks=true, urlcolor=\scolor, linkcolor=blue, citecolor=magenta, hyperindex=true, linktocpage=true]{hyperref}

\usepackage[explicit]{titlesec}

\def\N{{\mathcal{N}}}

\def\a{\alpha}

\def\({\left (}
\def\){\right )}
\def\[{\left[}
\def\]{\right]}

\def\be{\begin{equation}}
\def\ee{\end{equation}}
\newcommand{\bea}{\begin{eqnarray}}
\newcommand{\eea}{\end{eqnarray}}

\titleformat{\section}
{\gdef\sectionlabel{}
	\Large\bfseries\scshape}
{\gdef\sectionlabel{\thesection }}{0pt}
{\begin{tikzpicture}[remember picture]
	\draw (0.2, 0) node[right] {\color{\scolor} \textsf{#1}};
	\draw (0.0, 0) node[left, fill=\tcolor,circle] {\color{white} \textsf{\sectionlabel}};
	\end{tikzpicture}
}
\titlespacing*{\section}{0pt}{20pt}{5pt}
	\newcommand*\subsectionlabel{}
	\titleformat{\subsection}
	{\gdef\subsectionlabel{}
		\large\bfseries\scshape}
	{\gdef\subsectionlabel{\thesubsection  }}{0pt}
	{\begin{tikzpicture}[remember picture]
		\draw (0.15, 0) node[right] {\color{\scolor}\textsf{#1}}; 
		\draw[blue] (0.15, 0) node[left,fill=BurntOrange , circle]{ {\color{white} \textsf{\subsectionlabel}}};
		\end{tikzpicture}
	}
	\titlespacing*{\subsection}{5pt}{10pt}{0pt}

\pgfplotsset{compat=1.15}

\begin{document} 

	\allowdisplaybreaks
	
	\pagestyle{fancy}
	\renewcommand{\headrulewidth}{0pt}
	\fancyhead{}
	
	\fancyfoot{}
	\fancyfoot[C] {\textsf{\textbf{\thepage}}}
\begin{equation*}
	\begin{tikzpicture}
		\draw (0.5\textwidth, -3) node[text width = \textwidth]{ \huge \textsf{\textbf{Nonlinear response of the chiral magnetic effect in the D3/D7 holographic model}} };
		\draw[very thick, color=\scolor] (0.0\textwidth,-4) -- (\textwidth,-4);
	\end{tikzpicture}
\end{equation*}

\begin{equation*}
\begin{tikzpicture}
\draw (0.5\textwidth,0.2) node[text width=\textwidth] {\large \color{black}  \textsf{Masataka Matsumoto}$^{\color{\tcolor} \textsf{a,b}}$ \textsf{Mirmani Mirjalali}$^{\color{\tcolor} \textsf{c}}$ \textsf{Ali Vahedi}$^{\color{\tcolor} \textsf{d}}$};
\draw (0.5\textwidth, -0.5) node[text width=\textwidth] {\small $^{\color{\tcolor} \textsf{a}}$ \textsf{Department of Physics, Chuo University, 1-13-27 Kasuga, Bunkyo-ku, Tokyo 112-8551, Japan}};
\draw (0.5\textwidth,-1) node[text width=\textwidth] {\small $^{\color{\tcolor} \textsf{b}}$ \textsf{Department of Physics, Shanghai University, 99 Shangda Road, Shanghai 200444, China}};
\draw (0.5\textwidth, -1.5) node[text width=\textwidth] {\small $^{\color{\tcolor} \textsf{c}}$ \textsf{Faculty of Physics, Shahrood University of Technology, P.O. Box 3619995161, Shahrood, Iran}};
\draw (0.5\textwidth, -2.0) node[text width=\textwidth] {\small $^{\color{\tcolor} \textsf{d}}$ \textsf{Department of Astronomy and High Energy Physics, Kharazmi University, 15719-14911,
Tehran, Iran}};
\end{tikzpicture}
\end{equation*}

\begin{equation*}
	\begin{tikzpicture}
	\draw (0, -13.1) node[right, text width=\paperwidth] {\texttt{}};
	\draw (\textwidth, -13.1) node[left] {\textsf{\today}};
	\end{tikzpicture}
\end{equation*}
\begin{equation*}
	\begin{tikzpicture}
	\draw[very thick, color=\scolor] (0.0\textwidth, -5.75) -- (0.99\textwidth, -5.75);
	\draw (0.12\textwidth, -6.25) node[left] {\color{\scolor}  \textsf{\textbf{Abstract:}}};
	\draw (0.53\textwidth, -6) node[below, text width=0.8\textwidth, text justified] 
    {We investigate the nonlinear response of the chiral magnetic current to an external magnetic field in a holographic setup. Using the D3/D7 brane system, where the chiral magnetic effect (CME) can be realized by considering rotating probe D7-branes, corresponding to introducing an axial chemical potential, we analyze the current–magnetic field relation beyond the linear regime. 
    Focusing on the critical region between the insulating phase and the CME phase, we find that the chiral magnetic current exhibits a multi-valued dependence on the magnetic field,  which originates from the discrete self-similar structure emerging near the critical embedding characterized by a conical singularity of the probe D7-brane.
    We further study the dynamical stability of the insulating phase near the transition point, and show that the presence of both an axial chemical potential and an external magnetic field cooperatively stabilize the system. 
    We also discuss the thermodynamic stability of the multiple solutions by computing the thermodynamic potential.
    Our results clarify the interplay between the axial chemical potential and the magnetic field in determining the phase structure and stability of the system, and provide a new perspective on phase transitions associated with the multi-valued structure.
    };
	\end{tikzpicture}
\end{equation*}

\tableofcontents

\begin{equation*}
	\begin{tikzpicture}
	\draw[very thick, color=\scolor] (0.0\textwidth, -5.75) -- (0.99\textwidth, -5.75);
	\end{tikzpicture}
\end{equation*}

\section{Introduction}

Chiral magnetic effect (CME) is a remarkable transport phenomenon in which an electric current is induced along an external magnetic field in the presence of a chiral imbalance \cite{Fukushima:2008xe}. Originating from the quantum chiral anomaly, the CME has attracted considerable attention in a wide range of physical contexts, including relativistic heavy-ion collisions, condensed matter systems such as Weyl semimetals, and strongly coupled gauge theories. In particular, holographic approaches based on AdS/CFT correspondence \cite{Maldacena:1998,Gubser:1998,Witten:1998} have provided a powerful framework to study the CME beyond perturbation theory and at strong coupling.

Among various holographic realizations, the D3/D7 brane system offers a well-controlled setup in which flavor degrees of freedom are introduced into $\mathcal{N} = 4$ super Yang–Mills theory \cite{Karch:2002sh}. In this model, the CME can be realized by turning on an axial chemical potential, and the emergence of the chiral magnetic current has been studied in detail \cite{Hoyos:2011us}. Previous works have clarified the conditions under which the CME phase appears, as well as its interpretation from both the gravity side and the dual field theory perspective. A notable feature of the D3/D7 model is the existence of distinct phases, including an insulating phase without current and a CME phase characterized by a finite chiral magnetic current.

While much effort has been devoted to establishing the existence and qualitative features of the CME in this setup \cite{Hoyos:2011us}, its nonlinear response to external magnetic fields associated with the vicinity of the critical region has not been fully explored. This is particularly intriguing in light of the fact that the D3/D7 model exhibits highly nontrivial nonlinear current responses to external electric fields \cite{Karch:2007pd}, which have been extensively studied and shown to display rich phenomena such as multi-valued currents and phase transitions \cite{Nakamura:2010zd,Nakamura:2012ae,Ali-Akbari:2013hba}. From this perspective, it is natural to ask how the chiral magnetic current responds to strong magnetic fields beyond the linear regime, and whether similarly characteristic nonlinear behaviors arise in the CME.

In this paper, we systematically investigate the nonlinear response of the chiral magnetic current to an external magnetic field in the D3/D7 model. 
In particular, focusing on the vicinity of the critical region separating the insulating phase and the CME phase, we find that the current–magnetic field relation becomes multi-valued.
This multi-valued structure is associated with the discrete self-similarity around the critical region characterized by a conical singularity of the probe D7-brane configuration. We characterize the near-critical behavior by extracting the scaling law and the discrete self-similar exponents.
Furthermore, we analyze the dynamical stability of the insulating phase near the transition point and demonstrate that the magnetic field and the axial chemical potential can cooperatively stabilize the system. 
We also discuss the thermodynamic stability by computing the thermodynamic potential, corresponding to the renormalized on-shell action, and examine which branch is thermodynamically favored.
Our results reveal the critical behavior underlying the nonlinear response of the CME and provide a new perspective on phase transitions associated with the multi-valued structure.

The paper is organized as follows.
In section \ref{sec:review}, we briefly review a realization of the CME in the D3/D7 model.
In section \ref{sec:zeroT}, we show the current-magnetic field relation at zero temperature and discuss a nonlinear multi-valued behavior. Subsequently, we investigate the scaling law and the discrete self-similar exponents in the vicinity of the critical regime. We also investigate the dynamical stability of the multi-valued solutions by focusing on the meson spectrum in the insulator phase.
Additionally, we discuss the thermodynamic stability by computing the thermodynamic potential for each phase.
In section \ref{sec:finiteT}, we also show the results in the case of finite temperature.
Section \ref{sec:conclusion} is devoted to the conclusions and discussions.

\section{Review: Chiral magnetic effect from the probe brane model} \label{sec:review}
In this section, we briefly review the D3/D7 probe brane model \cite{Karch:2002sh}, which is well-known as one of top-down holographic models.
For our purposes, we consider a stack of $N_f$ D7-branes and a stack of $N_c$ D3 color branes.
In the limit of $N_c\to\infty$ and large 't$\,$Hooft coupling $\lambda = 4 \pi g_{s} N_{c}$ with $g_{s}N_{c}$ fixed, where $g_{s}$ is the string coupling, the background D3-branes are replaced by the supergravity solution which forms AdS$_5 \times S^5$ as a near horizon geometry. The AdS/CFT correspondence \cite{Maldacena:1998} claims that the system is dual to $\mathcal{N}=4$ supersymmetric $SU(N_c)$ Yang-Mills theory in $(3+1)$ dimensions, with the Yang-Mills coupling $g_{\rm YM}^{2} = 4 \pi g_{s}$. The near extremal solution of D3-branes, dual to a system at finite temperature, leads to a Schwarzschild-AdS geometry, which is given by the unit of AdS radius $L=1$
	\bea \label{metric0}
    \dd s^2&=&g_{tt}\dd t^2+g_{xx}\dd\vec{x}^2 + g_{uu} \dd u^2+g_{\theta\theta} \dd \theta^2+g_{\phi\phi} \dd\phi^2+g_{SS} \dd {\rm \Omega}_3 ^2,
	\eea 
	where
	\begin{equation} 
	\begin{split}
	g_{tt} &=-\frac{f(u)}{u^2}, \quad g_{xx}=\frac{1}{u^2},\quad g_{uu}=\frac{1}{u^2 f(u)}, \\
	g_{\theta\theta}&=1,\quad g_{\phi\phi}=\sin^2\theta,\quad g_{SS}= \cos^2\theta.
	\end{split}
	\end{equation}
		In the above, $f(u)=\left(1-u^4/u_h^4\right)$
	%  \bea
	%  b_h(u)=\left(1-u^4/u_h^4\right),
	%  \eea
	and $\dd{\rm \Omega}_3^2$ is the metric of $S^3$ with a unit radius. In this choice of coordinates, the boundary of the AdS is located at $u\to 0$ and the field theory lives in the $(3+1)$ dimensional Minkowski spacetime with coordinates $(t,\vec{x})$. The $u_h$ indicates the location of the black hole horizon and is related to the Hawking temperature, which corresponds to the heat bath temperature in the dual field theory, by
	\bea  \label{HawkT}
	T=\frac{1}{\pi u_h}.
	\eea
	At the $u_h\to \infty$ the Eq.~\eqref{metric0} becomes AdS$_5\times S^5$, corresponding to the system at zero temperature.
    
Consider the intersection of a stack of $N_f$ D7-branes with a stack of $N_{c}$ D3 branes, which are embedded in ten-dimensional spacetime as Table~\ref{tab:1}:
\begin{table}[H]
	\centering
	\begin{tabular}{|c|c|c|c|c|c|c|}
		\hline
		&$t$&$\vec{x}$& $u$ &$S^3$&$\theta$&$\phi$\\
		\hline 
		D3&$\times $& $\times $ &$ $ & $ $&&\\
		\hline
		D7&$\times $ & $\times $ & $\times $&$\times$&&\\
		\hline
	\end{tabular}
	\caption{\label{tab:1} The embeddings of the D3-branes and D7-branes in ten-dimensional spacetime.}
\end{table} 
\noindent The probe limit corresponds to ${N_f}\ll {N_c}$ and the dynamics of probe D7-branes is governed by the Dirac-Born-Infeld (DBI) action and Wess-Zumino (WZ) term 
\be\label{D7Action}
S_{D7} = S_{\rm DBI} + S_{\rm WZ},
\ee
\bea\label{Action0}
S_{\rm DBI} & = &- N_f T_{D7} \int d^8\sigma \sqrt{-\textrm{det}\left( g_{ab}^{D7} + \left( 2\pi \a'\right) F_{ab}\right)}, \\
S_{\rm WZ} & = & + \frac{1}{2} N_f T_{D7} \left( 2\pi\a'\right)^2\int P[C_4] \wedge F \wedge F,\label{Action1}
\eea
where $T_{D7} = {g_s^{-1} \a'^{-4}}{(2\pi)^{-7}}$ is the D7-brane tension with the string length squared $\alpha'$, $\sigma^a~(a = 0,\cdots7)$ are the worldvolume coordinates, 
% and $P[]$ stands for the pullback of the background field to the $D7$ probe branes 
$g_{ab}^{D7}$ is the induced metric on the probe D7-branes or pullback of the background metric to the D7-branes, $F_{ab} = \partial_{a} A_{b} - \partial_{b} A_{a}$ is the worldvolume field strength for the $U(1)$ gauge field $A_{a}$ living on the D7-branes, and $P[C_4]$ is the pullback of the Ramond-Ramond four-form to the probe D7-branes.
The induced metric is explicitly written as
\begin{equation}
    g_{ab}^{D7} = \frac{\partial X^{\mu}}{\partial \sigma^{a}}\frac{\partial X^{\nu}}{\partial \sigma^{b}} g_{\mu\nu},
\end{equation}
where $X^{\mu}~(\mu=0,\cdots,9)$ denotes the target space coordinates and $g_{\mu\nu}$ is the background metric given by \eqref{metric0}. In our conventions, the Greek letters ($\mu,\nu$) and Latin letters ($a,b$) denote the indices for the ten-dimensional background coordinates and eight-dimensional worldvolume coordinates, respectively.
The Ramond-Ramond four-form is explicitly given by
\be\label{eq:fourform0}
C_4 = g_{xx}^2 \, \textrm{vol}_{\mathbb{R}^{1,3}} - g_{SS}^2 \, d\phi \wedge \textrm{vol}_{S^3}.
\ee
The two scalar fields $(\theta, \phi)$ determine the configuration of the D7-branes in ten-dimensional spacetime, and are dual to the operators ${{\mathcal{O}}_{m}}$ and ${{\mathcal{O}}_{\phi}}$, respectively. 
More specifically, their asymptotic values correspond to the modulus and phase of the complex mass as discussed later.
In the following, we consider their geometrical roles in bulk and the corresponding physical meaning in the dual field theory.

With the D-brane intersections as shown in Table~\ref{tab:1}, where the probe D7-branes extended along AdS$_5\times S^3$, the system has a rotational symmetry in two dimensional plane, parametrized by $(\theta,\phi)$ coordinates, if the probe D7-branes are located at the position overlapping the D3-branes, namely $\theta(u)=0$. This rotational symmetry, on the other hand, is broken if we consider the non-trivial D7-brane configuration $\theta(u)\neq 0$. The rotational symmetry $SO(2)$ in the bulk theory corresponds to the global axial (chiral) $U(1)_A$ symmetry in the dual field theory. The chiral symmetry in the D3/D7 system and its spontaneous breaking have been extensively investigated in the previous studies \cite{Babington:2003vm,Filev:2007gb,Filev:2007qu,Albash:2007bk,Albash:2007bq,Erdmenger:2007bn,Zayakin:2008cy,Filev:2009xp,Evans:2011tk}.

Regarding the ansatz for $\phi$, we consider the form of $\phi(t,u)=\omega t+\varphi(u)$ where $\omega$ is the angular frequency of the rotating D7-branes, introducing a time-dependent phase of the complex mass as $|m|\,e^{i\omega t}$.
As discussed in \cite{Hoyos:2011us}, the angular frequency $\omega$ is a bulk dual of axial chemical potential $\mu_5$ in the boundary field theory as $\omega=2\mu_5$.

To study the chiral magnetic effect we also introduce a constant magnetic field $B$ and $U(1)_{V}$ current along $z$ direction on the probe D7-branes.
In total, our ansatz for the fields are
\begin{equation}
    \theta = \theta(u), \quad \phi = \omega t + \varphi(u), \quad (2\pi\alpha')A_{y} = B x, \quad (2\pi\alpha')A_{z} = A_{z}(u). 
\end{equation}
Here, a factor of $(2\pi\alpha')$ is absorbed into the gauge fields.
With these assumptions the DBI action \eqref{Action0} and WZ term \eqref{Action1} becomes
\begin{align}\label{actionden}
\tilde{S}_{DBI} &= -\N \int \dd u \, \sqrt{g_{xx}^{3}g_{SS}^{3}} \, \sqrt{ 1 + \frac{B^2}{g_{xx}^2}} \sqrt{-(g_{tt}+g_{\phi\phi}\omega^{2})\left(g_{uu}+\theta'^2 + \frac{A_z'^2}{g_{xx}} \right) -  g_{tt} g_{\phi\phi}\varphi'^2}  , \nonumber\\
\tilde{S}_{WZ} &= -\N B \omega \int \dd u \, g_{SS}^2 A_z' ,
\end{align} 
where primes denote the derivative with respect to $u$ coordinate. We also define the action density as $\tilde{S}={S}/{\rm Vol_{\mathbb{R}^{1,3}}}$.
The overall factor is defined as 
\begin{equation}
    \N=N_f T_{D7} \, (2\pi^2) = \frac{\lambda N_f N_c}{(2\pi)^4},
\end{equation}
with the relation $4 \pi g_{s} N_{c}\alpha'^{2} = 1$.
We have three equations of motion for $\{ \theta,\varphi,A_{z} \}$, but since the actions do not depend on $A_z(u)$ and $\varphi(u)$, we have two constants of motion,
%\footnote{we pick the same notations as \cite{Hoyos:2011us} for the constants of the motion.}
  \begin{eqnarray}\label{eq:alpha}
  \frac{\delta \tilde{S}_{D7}}{\delta \varphi'}&=& +\N \frac{\sqrt{g_{xx}^{3}g_{SS}^{3}} \, \sqrt{1 + \frac{B^2}{g_{xx}^2}} g_{tt}g_{\phi\phi}\varphi'}{\sqrt{-(g_{tt}+g_{\phi\phi}\omega^{2})\left(g_{uu}+\theta'^2 + \frac{A_z'^2}{g_{xx}} \right)- \, g_{tt} g_{\phi\phi}\varphi'^2}} \equiv \alpha ,\\
  \frac{\delta \tilde{S}_{D7}}{\delta A'_z}&=&+\N \frac{\sqrt{g_{xx}^{3}g_{SS}^{3}} \, \sqrt{1 + \frac{B^2}{g_{xx}^2}} \,\frac{(g_{tt}+g_{\phi\phi}\omega^{2})A_{z}'}{g_{xx}}}{\sqrt{-(g_{tt}+g_{\phi\phi}\omega^{2})\left(g_{uu}+\theta'^2 + \frac{A_z'^2}{g_{xx}} \right)-  g_{tt} g_{\phi\phi}\varphi'^2}} - \N B \omega g_{SS}^{2} \equiv\beta. \label{eq:beta}
  \end{eqnarray}
Combining them, we obtain the relation
\begin{equation}
    (g_{tt}+g_{\phi\phi}\omega^{2}) \alpha A_{z}' - g_{tt}g_{\phi\phi}g_{xx} \left(\beta + \N B \omega g_{SS}^{2} \right)\varphi' =0, \label{eq:relation}
\end{equation}
which implies that the two constants of motion are not independent and related to each other for arbitrary $u$. With this relation, we can eliminate $A_{z}'$ in \eqref{eq:alpha} and we obtain
\begin{equation}
    \N \frac{\sqrt{g_{xx}^{3}g_{SS}^{3}} \, \sqrt{1 + \frac{B^2}{g_{xx}^2}} g_{tt}g_{\phi\phi}\varphi'}{\sqrt{-(g_{tt}+g_{\phi\phi}\omega^{2})\left(g_{uu}+\theta'^2 \right)- \left( 1+ \frac{g_{tt} g_{\phi\phi} g_{xx}(\beta +\N B \omega g_{SS}^{2})^{2}}{ (g_{tt} + g_{\phi\phi}\omega^{2})\alpha^{2}} \right) g_{tt}g_{\phi\phi}\varphi'^2}} = \alpha. \label{eq:alpha2}
\end{equation}
Here the factor of $g_{tt} + g_{\phi\phi} \omega^{2}$ can become zero at a certain value of $u$, because $g_{tt}$ is zero at the black hole horizon and $-\infty$ at the boundary whereas $g_{\phi\phi} \omega^{2}$ is always positive. We denote this position as $u=u_{*}$ with $u_{h}>u_{*}>0$. In the limit of $u\to u_{*}$, \eqref{eq:relation} leads to $\beta + \N B\omega g_{SS}^{2}\to 0$ because $g_{tt}g_{\phi\phi}g_{xx}$ keeps finite at $u_{*}$.\footnote{Here, we also assume $\varphi'(u\to u_{*})\neq0$. 
If $\varphi'$ approaches zero in the limit of $u\to u_{*}$ along with $g_{tt} + g_{\phi\phi} \omega^{2}$, $\alpha$ also becomes zero from \eqref{eq:alpha2} and we do not consider this solution here.
} Note that the factor of $g_{tt} + g_{\phi\phi} \omega^{2}$ and $\beta + \N B\omega g_{SS}^{2}$ should approach zero in the same power. Thus, we derive the form of $\alpha$, which is independent of $\{\varphi,A_z \}$, from \eqref{eq:alpha2} as
\begin{equation}
    \alpha^{2} = \N^{2}(-g_{tt}g_{\phi\phi}g_{xx}^{3}g_{SS}^{3}) \, \left( 1 + \frac{B^2}{g_{xx}^2} \right) \Bigg|_{u_{*}},
\end{equation}
where each metric component is evaluated at $u=u_{*}$. The location of $u_{*}$ is referred to as a world-volume horizon (or effective horizon) since it forms a causal boundary for fluctuations on the D7-branes \cite{Seiberg:1999vs,Kim:2011qh}. In summary, the constants of motion are determined without obtaining the explicit forms of $\{\varphi', A_{z}' \}$ via 
\begin{align}
    &(g_{tt} + g_{\phi\phi} \omega^{2}) \big|_{u_{*}} = 0, \label{eq:ustar1}\\
    &\alpha^{2} = \N^{2}(-g_{tt}g_{\phi\phi}g_{xx}g_{SS}^{3}) \, \left( g_{xx}^{2} + B^2 \right) \big|_{u_{*}}, \label{eq:ustar2}\\
    &\beta = -\N B \omega g_{SS}^{2} \big|_{u_{*}},\label{eq:ustar3}
\end{align}
where the first equation determines the location of $u_{*}$.

Now we perform the Legendre transformation with respect to the action as
  \be
  \hat{S}_{D7}=\tilde{S}_{D7}- \int \dd u \left(\alpha \varphi'+\beta A_z'\right),
  \ee
and substitute the solutions
\begin{eqnarray}
    \varphi'^{2} &=& \frac{(g_{tt} + g_{\phi\phi} \omega^{2})^{2}(g_{uu} + \theta'^{2}) \frac{\alpha^{2}/ \N^{2}}{g_{tt}^{2}g_{\phi\phi}^{2}} }{ - g_{xx}^{3} g_{SS}^{3} (g_{tt} + g_{\phi\phi} \omega^{2}) \left(  1+ \frac{B^{2}}{g_{xx}^{2}}+\frac{\alpha^{2}/ \N^{2}}{g_{tt}g_{xx}^{3}g_{\phi\phi}g_{SS}^{3}} \right) -g_{xx} \left(\frac{\beta}{\N} + B\omega g_{SS}^{2} \right)^{2}}, \\
    A_z'^{2} &=& \frac{(g_{uu} + \theta'^{2}) g_{xx}^{2} \left(\frac{\beta}{\N}+ B\omega g_{SS}^{2} \right)^{2} }{- g_{xx}^{3} g_{SS}^{3} (g_{tt} + g_{\phi\phi} \omega^{2}) \left(  1+ \frac{B^{2}}{g_{xx}^{2}}+\frac{\alpha^{2}/ \N^{2}}{g_{tt}g_{xx}^{3}g_{\phi\phi}g_{SS}^{3}} \right) -g_{xx} \left(\frac{\beta}{\N} + B\omega g_{SS}^{2} \right)^{2}}.
\end{eqnarray}
Thus, we could solve the equations for $\varphi$ and $A_z$ in terms of the constants of motion and put them back into the action, and the transformed action becomes
\bea
\label{eq:Shat}
\hat{S}_{D7} = -\N \int \dd u \sqrt{g_{uu} + \theta'^2}\sqrt{-g_{xx}^3 g_{SS}^3 \left(g_{tt}+g_{\phi\phi} \omega^2\right)\left(1 + \frac{B^2}{g_{xx}^2} + \frac{\alpha^2/\N^2}{g_{tt} g_{xx}^3 g_{\phi\phi} g_{SS}^3}\right)- g_{xx} \left( \frac{\beta}{\N} + B \omega g_{SS}^2\right)^2},
\eea
which contains only the field $\theta$. Note that demanding the reality condition for this action, we obtain the same equations as Eqs.~\eqref{eq:ustar1}-\eqref{eq:ustar3} as discussed in \cite{Hoyos:2011us}.

To associate the bulk fields to the boundary quantities, we expand the bulk fields in the vicinity of the boundary as
\begin{eqnarray}
    \theta(u) &=& c_{0} u + c_{2} u^{3} - \frac{c_{0}}{2}\omega^{2} u^{3} \log u + \cdots,  \label{eq:asym1}\\
    \phi(t,u) &=& \omega t + \frac{\alpha}{2 {\cal{N}} c_{0}^{2}} u^{2} +\cdots, \\
    A_{z}(u) &=& c_{z} + \frac{1}{2}\left( \frac{\beta}{\N} + B \omega \right) u^{2} + \cdots.
\end{eqnarray}
According to the holographic dictionary \cite{Hoyos:2011us}, we have
\begin{align}
\begin{split}
    & |m|=\frac{c_{0}}{2\pi \alpha'},\quad \expval{\mathcal{O}_{m}} = -2\pi \alpha' \N \left( 2 c_{2} - \frac{1}{3} c_{0}^{3} +\frac{1}{2}\omega^{2} c_{0} + \frac{1}{2}\omega^{2}c_{0}\log c_{0}^{2} \right) \\
    &\expval{\mathcal{O}_{\phi}} = \alpha ,\quad \expval{J^z} =-2\pi \alpha' \beta. \label{eq:dictionary}
\end{split}
\end{align}
As mentioned above, we also have the axial chemical potential and boundary magnetic field as $\mu_{5} = \omega /2 $ and $B/(2\pi \alpha')$, respectively. Without loss of generality, we set the source for $\expval{J^{z}}$ to zero, $c_{z}=0$.

What remains is to determine the D7-brane embedding, that is, to solve the equation of motion for $\theta(u)$ derived from \eqref{eq:Shat}. Since the equation is a nonlinear ordinary differential equation, we need to solve and find solutions numerically. Depending on the boundary condition for $\theta(u)$, there are three types of embedding: black hole embedding, Minkowski embedding with a world-volume horizon, and Minkowski embedding without a world-volume horizon. For black hole embedding, the D7-brane falls into the black hole, and necessarily has the world-volume horizon \cite{Hoyos:2011us}. Thus, the Minkowski embedding with a world-volume horizon is equivalent to the black hole horizon at finite temperature. At zero temperature, however, it is possible that the world-volume horizon emerges on the D7-brane at $u=u_{*}$ in the AdS$_{5}\times S^{5}$ background geometry. To numerically find those solutions, we impose a certain value $\theta(u_{*})$ and the condition for $\theta'(u_{*})$ derived from the equation of motion as the boundary condition. Given a finite $\theta(u_{*})$ and external parameters of $B$ and $\omega$, we can determine the values of $\alpha$ and $\beta$ from \eqref{eq:ustar2} and \eqref{eq:ustar3}, corresponding to some finite values of the pseudo-scalar condensate and current density along $z$ direction. Once we find a solution $\theta(u)$, we can read off the quark mass $|m|$ from the asymptotic behavior \eqref{eq:asym1}. In this work, we will use the quark mass as the scale of other quantities.  This is how we compute the current density $\expval{J^{z}}$ as a function of $B$ with $\omega$ and $T$ fixed.

For Minkowski embedding without a world-volume horizon, the D7-brane does not reach the black hole horizon and the world-volume horizon is also not emerged, corresponding the solutions with $\alpha = \beta =0$. As discussed in \cite{Hoyos:2011us}, the solutions without a world-volume horizon correspond to states with no CME and no spontaneous breaking of CT, while those with a world-volume horizon correspond to states with a CME and spontaneous breaking of CT. 
The condensate $\expval{{\cal{O}}_{\phi}}$ serves as the order parameter for the spontaneous breaking of the CT symmetry.
To numerically find those solutions, we impose $\theta(u_{\rm max}) = \pi/2$ at $u_{*}>u_{\rm max}>0$ and $(\sin \theta(u_{\rm max})/u_{\rm max})'=0$ to avoid a world-volume horizon and conical singularity. The quark mass can be read off from \eqref{eq:asym1} again.

\begin{figure}
    \centering
    \includegraphics[width=0.7\linewidth]{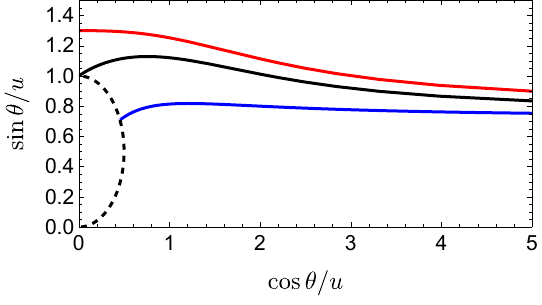}
    \caption{The typical solutions of the Minkowski embedding without a world-volume horizon (red), critical embedding (black), and Minkowski embedding with a world-volume horizon (blue). We here choose $B=\omega=1$ for simplicity. The dashed curve shows the location of the world-volume horizon.}
    \label{fig:embedding}
\end{figure}

The critical solution which connects the Minkowski embedding with and without a world-volume horizon is referred to as critical embedding. In figure \ref{fig:embedding}, we show the typical solutions of the Minkowski embedding without a world-volume horizon, critical embedding, and Minkowski embedding with a world-volume horizon, respectively. In this paper, we particularly focus on the solutions in the vicinity of the critical embedding, exhibiting the non-trivial multi-valued behaviors as discussed in the following section.

%%%%%%%%%%%%
\section{Zero temperature}\label{sec:zeroT}
\subsection{Current and condensate}
\begin{figure}[tbp]
    \centering
    \includegraphics[width=0.8\linewidth]{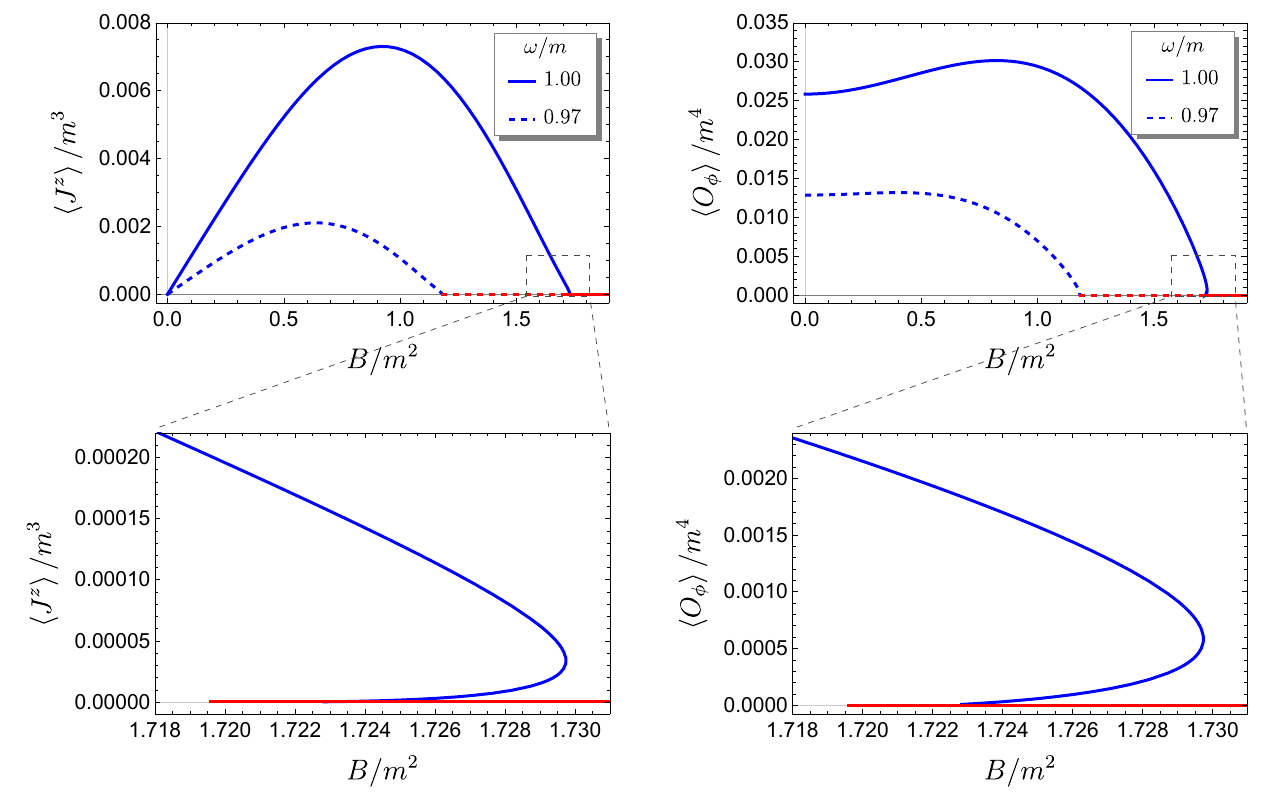}
    \caption{Upper panel: $\expval{J^{z}}/m^{3}$ and $\expval{\mathcal{O}_{\phi}}/m^{4}$ as a function of $B/m^{2}$ for $\omega /m = 1.00$ (solid) and $0.97$ (dashed). Two types of solutions, the Minkowski embedding with and without a world-volume horizon, are distinguished by blue and red colored lines, respectively. Lower panel: The zooming in to the region where $\expval{J^{z}}/m^{3}$ and $\expval{\mathcal{O}_{\phi}}/m^{4}$ become zero.}
    \label{fig:BJplotT0}
\end{figure}

In this section, we focus on the zero temperature case. At zero temperature ($u_{h}\to \infty$), the vector current along the $z$ direction and the pseudo-scalar mesons condensate are simply written as
\begin{align}
    \expval{J^{z}} &= - (2 \pi \alpha') \beta = (2 \pi \alpha') \N B \omega \cos^{4} \theta(u_{*}) , \\
    \expval{\mathcal{O}_{\phi}} &= \alpha = \N\frac{\sqrt{1+B^{2} u_{*}^{4}}}{u_{*}^{4}}\sin\theta(u_{*})\cos^{3}\theta(u_{*}).
\end{align}
where $u_{*}$ is determined by the equation $-1+ \omega^{2}u_{*}^{2}\sin^{2} \theta(u_{*})=0$.
In figure \ref{fig:BJplotT0}, we show $\expval{J^{z}}$ and $\expval{\mathcal{O}_{\phi}}$ as a function of $B$ for two values of $\omega$. Hereafter, we set ${\cal{N}}=1$ and $2\pi\alpha' =1$ in the numerical results, and all physical quantities are normalized by the quark mass $m$\footnote{For simplicity, we now write $m$ instead of $\abs{m}$.}. The upper panel shows that the current initially increases, then decreases, and eventually approaches zero as the magnetic field increases. The lower panel provides a zoomed-in view of the region where both the current and the condensate vanish, corresponding to the absence of the CME.

As discussed in \cite{Hoyos:2011us}, the current density and pseudo-scalar condensate increase for smaller $B$ while decrease and finally vanish for larger $B$. In the dual field theory, the magnetic field induces the chiral symmetry breaking in our system and extinguishes the CME by restoring the CT symmetry.
In the bulk picture, this corresponds to the transition between the Minkowski embedding with and without a world-volume horizon. Interestingly, if we zoom in to the region of this transition as shown in the lower panel in figure \ref{fig:BJplotT0}, we see that $\expval{J^{z}}$ and $\expval{\mathcal{O}_{\phi}}$ do not monotonically approach zero, but they turn back to a certain value of $B$. In the D3/D7 model, this characteristic behavior is expected near the critical embedding, namely in the vicinity of the transition between the Minkowski embedding with and without a world-volume horizon, as mentioned below. Note that this folding structure is expected to repeat infinitely if we zoom in closer to the critical value of $B$. However, since the periodicity of this structure appears logarithmically, this behavior has not been closely illustrated in \cite{Hoyos:2011us}.

\begin{figure}
    \centering
    \includegraphics[width=0.75\linewidth]{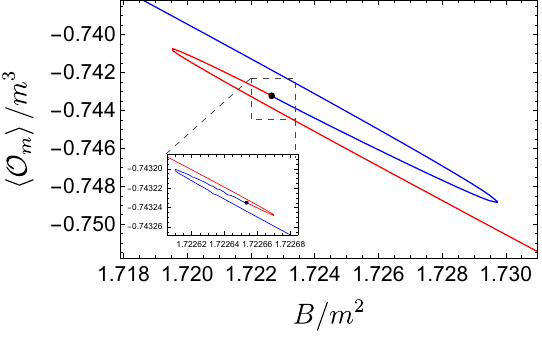}
    \caption{$\expval{{\cal{O}}_{m}}/m^{3}$ as a function of $B/m^{2}$ for $\omega /m = 1.00$. Two types of solutions, the Minkowski embedding with and without a world-volume horizon, are distinguished by blue and red colored lines, respectively. The black dot denotes the location of the critical embedding. The inset shows the repeated spiral structure in the vicinity of the critical embedding. The apparent zigzag structure near the critical embedding is a numerical artifact caused by the limited accuracy in extracting $c_{2}$ from the numerical solutions.}
    \label{fig:BcplotT0}
\end{figure}

To see this structure clearly, we check the mass operator $\expval{{\cal{O}}_{m}}$ given by \eqref{eq:dictionary}, which is conjugate to the quark mass and corresponds to the chiral condensate in the dual field theory. 
In figure \ref{fig:BcplotT0}, we show $\expval{{\cal{O}}_{m}}$ as a function of $B$ for $\omega/m=1$ around the transition between the Minkowski embedding with and without a world-volume horizon.
As expected from the previous results with an external electric field \cite{Ishigaki:2021vyv} instead of the rotation of the brane, the similar spiral structure is observed in our setup. The middle point between these two embeddings is %referred to as 
the critical embedding, which has a conical singularity at the effective horizon \cite{Hashimoto:2015wpa}. Note that this is a common feature in the vicinity of the critical embedding in various setups of the probe brane systems \cite{Mateos:2006nu,Mateos:2007vn,Filev:2007gb,Albash:2006ew,Albash:2007bq,Erdmenger:2007bn,Frolov:2006tc,Ishigaki:2023dss,Berenguer:2025oct}.

\begin{figure}
    \centering
    \includegraphics[width=0.32\linewidth]{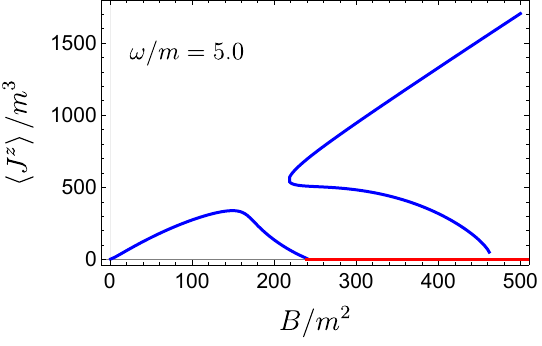}
    \includegraphics[width=0.32\linewidth]{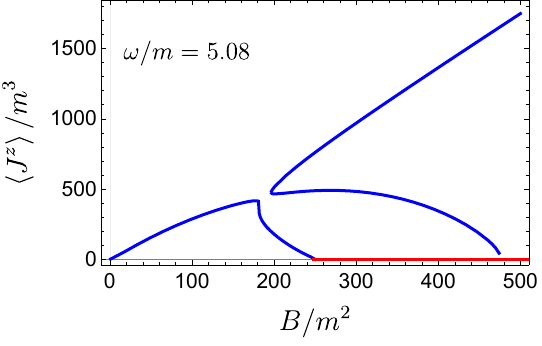}
    \includegraphics[width=0.32\linewidth]{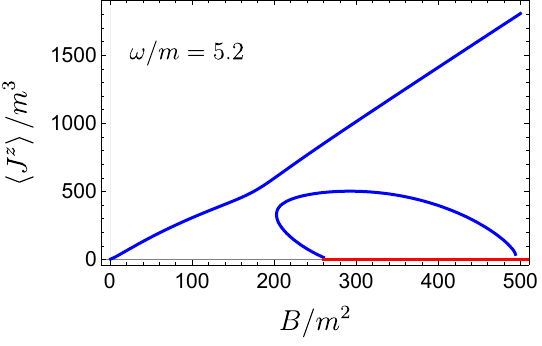}
    \caption{$\expval{J^{z}}/m^{3}$ as a function of $B/m^{2}$ for $\omega/m =5.0$ (left), $\omega/m=5.08$ (middle), and $\omega/m = 5.2$ (right). In each case, two distinct branches appear, and a reconnection between them occurs at a certain value of $\omega/m$. Two types of solutions, the Minkowski embedding with and without a world-volume horizon, are distinguished by blue and red colored lines, respectively.}
    \label{fig:BJplotL}
\end{figure}

So far, we have chosen smaller values of $\omega/m$. For larger values of $\omega/m$, we find that the $B$-$J^{z}$ curve is qualitatively changed as shown in figure \ref{fig:BJplotL}.
Firstly, we find that the current density becomes nonzero for arbitrarily large $B/m^{2}$.
%This is because in the bulk side there is no embedding transition and the only Minkowski embedding with a world-volume horizon is a possible solution for larger $\omega/m$.
This is because in the bulk side the Minkowski embedding with a world-volume horizon is always possible for any value of $B/m^{2}$, while the Minkowski embedding without a world-volume horizon appears for large $B/m^{2}$.
Secondly, two distinct branches appear in the $B$-$J^{z}$ plot. 
Interestingly, we find that the two branches reconnect at a certain value of $\omega/m$, as shown in the middle panel of figure \ref{fig:BJplotL}.
Note that the behavior is quite similar to the relation between the electric field and current density observed in the holographic Weyl semi-metal realized in the D3/D7 model \cite{Matsumoto:2024czp}. 
These nonlinear responses stem from the fact that we obtain multiple embedding solutions with respect to a given $B/m^{2}$ for this parameter region.

\subsection{Self-similarity near critical regime}
The multi-valued (or spiral) structure near the critical embedding as shown in figure \ref{fig:BJplotT0} and figure \ref{fig:BcplotT0} originates from the self-similarity around the conical singularity \cite{Mateos:2006nu,Mateos:2007vn,Frolov:2006tc,Ishigaki:2023dss,Berenguer:2025oct}. In this section, we analyze the discrete self-similar exponents that characterize the spiral structure.

For convenience, we use the coordinates
\begin{equation}
    R = \frac{\sin\theta(u)}{u}, \quad r = \frac{\cos \theta(u)}{u},
\end{equation}
and we have $u^{-2} = R^{2} +r^{2}$. The AdS boundary is now located at $r \to \infty$.
In this coordinate system, the background metric is given by
\begin{equation}
    \dd s^{2} = \left(r^{2} + R^{2} \right) \left(-\dd t^{2} + \dd \vec{x}^{2} \right) + \frac{1}{r^{2}+R^{2}} \left( \dd r^{2} + \dd R^{2} + r^{2} \dd {\rm \Omega}_{3}^{2} + R^{2} \dd \phi^{2} \right),
\end{equation}
and the location of the world-volume horizon $r=r_{*}$ is determined by
\begin{equation}
    \left( R(r_{*})- \frac{\omega}{2} \right)^{2} +r_{*}^{2} = \frac{\omega^{2}}{4}.
\end{equation}
In order to analyze near-critical solutions, we decompose the embedding function as $R(r) = R(r_{*}) +g(r) = \omega + g(r)$, and suppose that $g(r) = \varepsilon \tilde{g}(r/\varepsilon)$, $A_{z}(r) = \varepsilon \tilde{A}_{z}(r/\varepsilon)$, and $\varphi(r) =\varepsilon^{3/2} \tilde{\varphi}(r/\varepsilon)$.
Assuming $\varepsilon/\omega \ll 1$ and focusing on the region $0 \leq r \equiv \varepsilon \tilde{r}$, the DBI action becomes
\begin{eqnarray}
    \tilde{S}_{DBI} &\equiv&  {\cal{N}} \int \dd r \frac{\dd u}{\dd r} {\cal{L}} \nonumber \\
     &=&   -{\cal{N}} \int \dd r \,\,r^{3} \sqrt{\left( 1 + \frac{B^{2}}{(r^{2}+R^{2})^{2}} \right)\left[ \left(1- \frac{\omega^{2}R^{2}}{(r^{2}+R^{2})^{2}} \right)\left( 1 + A_{z}'^{2} + R'^{2} \right) + R^{2} \varphi'^{2} \right]}\nonumber\\
     &\sim& -\int \dd \tilde{r} \,\, \tilde{r}^{3} \sqrt{\tilde{g}\left(1+\tilde{A_{z}'^{2}}+\tilde{g}'^{2} \right)+\omega^{2}\tilde{\varphi}'^{2}},
\end{eqnarray}
in the leading order of $\varepsilon$. In the last expression, we omit the factor which does not affect the equations of motion for the fields.
The resulting action possesses a scaling symmetry: $\tilde{g} \to \tilde{g}(\mu \tilde{r})/\mu$, $\tilde{A}_{z} \to \tilde{A}_{z}(\mu \tilde{r})/\mu$, and $\tilde{\varphi} \to \tilde{\varphi}(\mu \tilde{r})/\mu^{3/2}$. Thus, once a set of solutions is obtained, one can generate other solutions through this scaling transformation for any positive $\mu$.

Now we consider the linear perturbations around the critical solutions as $g(r) = g_{c}(r) + \delta g(r)$, $A_{z}(r) = A_{z,c}(r)+ \delta A_{z}(r)$, and $\varphi(r) = \varphi_{c}(r)+\delta \varphi(r)$ with $A_{z,c}=\varphi_{c} =0$. The equations of motion for the perturbations are given by
\begin{equation}
    \delta g'' + \frac{7}{2r} \delta g ' + \frac{7}{2 r^{2}} \delta g =0, \quad \delta A_{z}'' + \frac{7}{2 r} \delta A_{z}' =0, \quad \delta \varphi'' + \frac{5}{2 r} \delta \varphi' =0.
\end{equation}
Then, the general solutions are 
\begin{eqnarray}
    \delta g (\tilde{r}) &=& C_{1} \tilde{r}^{-5/4}\sin \left( \frac{\sqrt{31}}{4}\log\tilde{r} \right) + C_{2} \tilde{r}^{-5/4}\cos\left( \frac{\sqrt{31}}{4}\log\tilde{r} \right) \nonumber \\
    \delta \tilde{A}_{z}(\tilde{r}) &=&C_{3} + C_{4} \tilde{r}^{-5/2},  \\
    \delta \tilde{\varphi}(\tilde{r}) &=& C_{5} + C_{6} \tilde{r}^{-3/2} \nonumber,
\end{eqnarray}
where the coefficients $C_{n}$ should be determined by imposing the explicit boundary conditions.
Note that the solution of $\delta g$ can be written as the form of $\delta g \sim \tilde{r}^{\nu_{\pm}}$ with the complex exponents {$\nu_{\pm} = (-5 \pm i\sqrt{31})/4$}.
These discrete self-similar exponents coincide with those for the effective horizon induced by an external electric field \cite{Ishigaki:2023dss}, suggesting a universal behavior determined solely by the local geometry of the collapsing $S^{3}$.
The scaling symmetry makes the coefficients transform as
\begin{equation}
 \begin{pmatrix}
  C_1 \\ C_2
 \end{pmatrix}
 \to \frac{1}{\mu^{9/4}}
 \begin{pmatrix}
  \cos\left(\frac{\sqrt{31}}{4}\log \mu\right) &
  - \sin\left(\frac{\sqrt{31}}{4}\log \mu\right) \\
  \sin\left(\frac{\sqrt{31}}{4}\log \mu\right) &
  \cos\left(\frac{\sqrt{31}}{4}\log \mu\right)
 \end{pmatrix}
  \begin{pmatrix}
  C_1 \\ C_2
 \end{pmatrix}
 ,
\end{equation}
\begin{equation}
 \begin{pmatrix}
  C_3 \\ C_4
 \end{pmatrix}
 \to 
 \begin{pmatrix}
  1/\mu & 0 \\
  0 & 1/\mu^{7/2}
 \end{pmatrix}
  \begin{pmatrix}
  C_3 \\ C_4
 \end{pmatrix}, \quad 
 \begin{pmatrix}
  C_5 \\ C_6
 \end{pmatrix}
 \to 
 \begin{pmatrix}
  1/\mu^{3/2} & 0 \\
  0 & 1/\mu^{3}
 \end{pmatrix}
  \begin{pmatrix}
  C_5 \\ C_6
 \end{pmatrix}
 .
\end{equation}
This result implies that the brane embeddings show discrete self-similarity with a period given
by $\mu = \mu_{2\pi}\equiv \exp(8\pi/\sqrt{31})$ as approaching the critical embedding. It yields the spiral behavior of the $B$-$\expval{{\cal{O}}_{m}}$ curves in the vicinity of the critical embedding as shown in figure \ref{fig:BcplotT0}.

\begin{figure}
    \centering
    \includegraphics[width=0.8\linewidth]{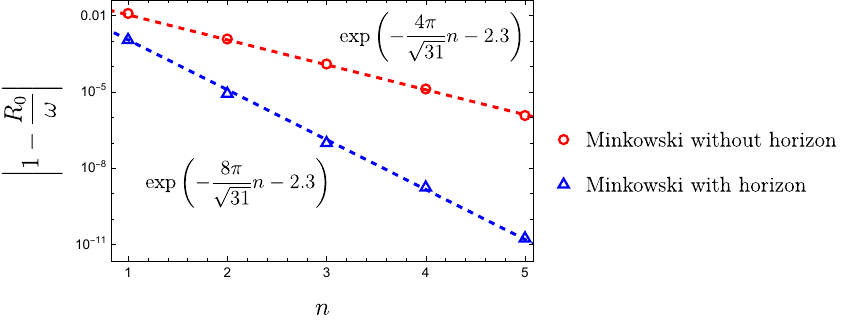}
    \caption{The values of the boundary condition $|1-R_{0}/\omega|$ at the $n$-th turning point of the $B$-$\expval{{\cal{O}}_{m}}$ curves. The limit $n\to \infty$ corresponds to the critical embedding. Each turning point is obtained from the numerical calculations. The dashed lines denote semi-analytical predictions for the period based on the discrete self-similarity, in which the values with $n \to 0$ are determined by the fitting. Note that $R_0 = R(r=0)$ for the Minkowski embedding without a world-volume horizon and $R_0 = R(r=r_{*})$ for the Minkowski embedding with a world-volume horizon.}
    \label{fig:dsi}
\end{figure}

To verify the discrete self-similarity in our numerical results, we consider the boundary conditions in the vicinity of the critical region. For the Minkowski embedding without a world-volume horizon, the brane embedding at the center $r=0$ behaves as 
\begin{equation}
    R(r) - \omega = \alpha_{0} + \alpha_{2}r^{2}+\cdots,
\end{equation}
where $\alpha_{0}$ and $\alpha_{2}$ are constant coefficients. Note that it contains only even powers in $r$ due to the symmetry under $r\to-r$.
In the limit of $r\to0$, the boundary condition for the Minkowski embedding without a world-volume horizon becomes
\begin{equation}
    R_{0} - \omega = \alpha_{0} \to \frac{\alpha_{0}}{\mu},
\end{equation}
where $R_{0}\equiv R(r=0)$ and the arrow indicates the rescaling of the boundary condition via the scaling transformation. This follows from $g(r=0) = \alpha_{0}$ in the present case.
For the Minkowski embedding with a world-volume horizon, on the other hand, the brane embedding near the world-volume horizon $r=r_{*}$ becomes
\begin{equation}
    R(r) - \left(\frac{\omega}{2}+\sqrt{\frac{\omega^{2}}{4}-r_{*}^{2}} \right) = \alpha_{1}(r-r_{*}) + \cdots,
\end{equation}
where $\alpha_{1}$ is a constant coefficient.
Thus, the boundary condition for the Minkowski embedding with a world-volume horizon can be written as
\begin{equation}
     \omega - R_{0}  \simeq \frac{r_{*}^{2}}{\omega} \to \frac{r_{*}^{2}}{\omega \mu^{2}},
\end{equation}
where, in this case, $R_{0} \equiv R(r_{*})$.
Note that under the scaling transformation, $g(r) \to g(\mu r)/\mu $, the zero of $g(r)$ is mapped as $r_{*} \to r_{*}'\equiv r_{*} / \mu$ since $g(r_{*})=0$ implies $g(\mu r_{*}') =0$.
These scalings of the boundary conditions leads to the turning points in the spiral structure at every half-period $\frac12 \log \mu_{2\pi}$. 
In figure $\ref{fig:dsi}$, we show the appearance of the $n$-th turning point and the corresponding boundary-condition values for the Minkowski embedding without and with a world-volume horizon. The dashed lines represent the semi-analytical predictions: $R_{0}-\omega \propto \mu_{2\pi}^{-n/2}$ and $\omega - R_{0} \propto \mu_{2\pi}^{-n}$ for the Minkowski embedding without and with a world-volume horizon, respectively.

\subsection{Meson spectrum}
Here, we discuss the meson spectrum in the Minkowski embedding without a world-volume horizon, corresponding to the normal mode for the fluctuations on the D7-branes. 
%For convenience, we use the coordinates
%\begin{equation}
%    R = \frac{\sin\theta(u)}{u}, \quad r = \frac{\cos \theta(u)}{u},
%\end{equation}
%and we have $u^{-2} = R^{2} +r^{2}$. The AdS boundary locates at $r \to \infty$ in this coordinate.
Since $\alpha = \beta = 0$ in the Minkowski embedding without a world-volume horizon, the only $R(r)$ is a non-trivial background solution. Then, we consider the fluctuations 
\begin{equation}
    R \to R(r) + \delta R(t,r), \quad \varphi \to 0 + \delta \varphi(t,r), \quad A_{z} \to 0 + \delta A_{z}(t,r),
\end{equation}
and expand the action up to the quadratic order in the fluctuations. Note that the fluctuations of the other gauge field components are decoupled to them. After we derive the equations of motion for the fluctuations, we perform the Fourier transformation with respect to $t$ as
\begin{equation}
    \delta R (t,r) = \int \frac{\dd \omega}{2\pi} \delta \tilde{R} (r) e^{-i k_{0} t}, \quad
    \delta \varphi (t,r) = \int \frac{\dd \omega}{2\pi} \delta \tilde{\varphi} (r) e^{-i k_{0} t}, \quad
    \delta A_{z} (t,r) = \int \frac{\dd \omega}{2\pi} \delta \tilde{A}_{z} (r) e^{-i k_{0} t}, 
\end{equation}
and obtain the eigenvalue equations for $\Phi\equiv\{\delta \tilde{R},\delta \tilde{\varphi}, \delta \tilde{A}_{z} \}$ with the frequency $k_{0}$.
The explicit forms of the equations are shown in Appendix \ref{sec:AppA}.

We consider the boundary conditions required to numerically solve the equations of motion. Since we employ the shooting method, we impose the boundary conditions at $r=0$, where the radius of $S^{3}$ vanishes and it becomes a regular singular point of the equations.
To avoid a singularity and obtain physical solutions, we impose the regularity condition, $\Phi'(r=0)=0$.  We also fix $\Phi(r=0)$ to specific values and read off the asymptotic behaviors of $\Phi(r=\infty)$ from the numerical solutions. Using the determinant method \cite{Kaminski:2009dh}, we determine the normal mode frequency $k_{0}$ by numerically solving the equations of motion with a linearly independent set of $\Phi(r=0)$. 

\begin{figure}
    \centering
    \includegraphics[width=1.\linewidth]{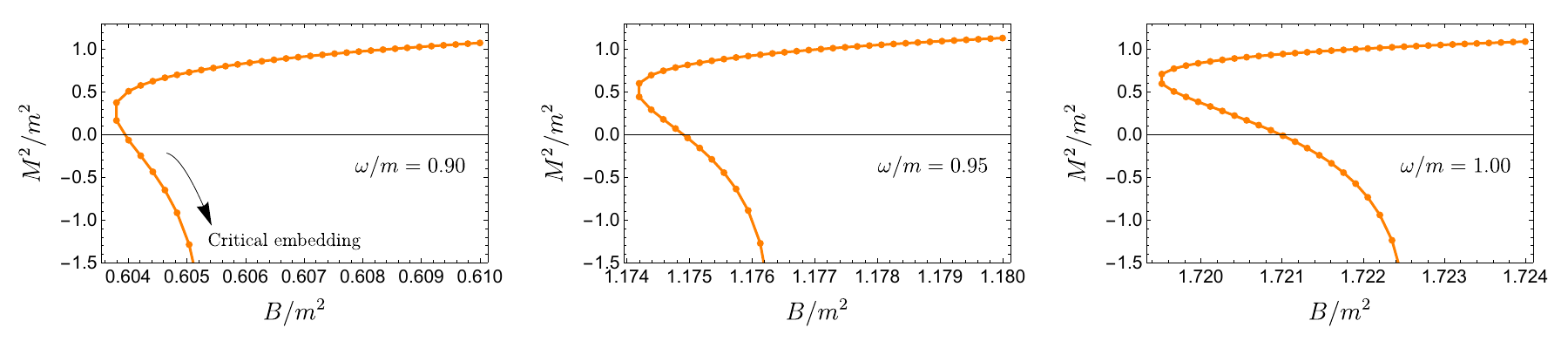}
    \caption{The plots of $M^{2}/m^{2} = (k_{0})^{2}/m^{2} $ as a function of $B/m^{2}$ for several values of $\omega/m$ in the Minkowski embedding without a world-volume horizon.}
    \label{fig:M2plot}
\end{figure}

For our purposes, we compute the mass square $M^{2}\equiv (k_{0})^{2}$ and show the results for several values of $\omega/m$ in figure \ref{fig:M2plot}.
Corresponding to the multiple solutions for a given $B/m^{2}$ as observed in figure \ref{fig:BcplotT0}, we obtain the multi-valued mass squared with respect to the magnetic field.
We find that the  solutions for a certain region of $B/m^{2}$ can be dynamically stable ($M^{2}>0$), while the solutions become tachyonic and unstable ($M^{2}<0$) as they approach the critical embedding.
Interestingly, the turning point does not necessarily correspond to the switch of the dynamical stability, but both of two distinct solutions for a certain $B$ can be stable.
Our results are distinct from the results with different setups in the D3/D7 model \cite{Mateos:2006nu,Mateos:2007vn,Ishigaki:2021vyv}, where one branch is always stable and the other is unstable.
%This is interpreted as indicating that the coupling of the fluctuations via $\omega$ and $B$ could stabilize the system, corresponding to the fact that effective mass of the meson bound state can be shifted in the presence of the axial chemical potential and the external magnetic field. 
This is interpreted as indicating that the coupling among fluctuation modes via $\omega$ and $B$ modifies the mass spectrum of the meson bound states, shifting $M^{2}$ to positive values and thereby stabilizing the system.
As evidenced by the plots, the onset of the dynamical instability approaches the turning point of the curve as $\omega/m$ decreases, as shown in figure \ref{fig:M2plot}.

\subsection{Thermodynamic potential}
We focus on the dynamical stability in the previous section.
In this section, we investigate the thermodynamic stability determined by the thermodynamic potential.
To do so, we assume that the thermodynamic potential can be holographically identified by the renormalized on-shell bulk action.
Since the on-shell action is divergent at the AdS boundary, we introduce the UV cutoff $u=\epsilon$ and its divergent part is given by 
\begin{equation}
    \tilde{S}_{D7} = {\cal{N}} \left[ -\frac{1}{4\epsilon^{4} }+\frac{c_{0}^{2}}{2 \epsilon^{2}} - \omega^{2} c_{0}^{2} \log \epsilon + \frac{B^{2}}{2}\log \epsilon + \cdots\right].
\end{equation}
To regulate the divergence of the action, we need to add the corresponding counterterms following holographic renormalization \cite{Karch:2005ms}.
The explicit counterterms are given by
\cite{OBannon:2008cmg,Hoyos:2011us}
\begin{eqnarray}
    L_{1} &=& {\cal{N}} \frac{1}{4 \epsilon^{4}}, \\
    L_{2} &=& {\cal{N}} \left( -\frac{c_{0}^{2}}{2 \epsilon^{2}} - c_{0} c_{2} +\frac12 \omega^{2} c_{0}^{2} \log \epsilon, \right) \\
    L_{3} &=& {\cal{N}} \frac{5}{12}c_{0}^{4}, \\
    L_{4} &=& {\cal{N}} \frac{1}{2}\omega^{2}c_{0}^{2} \log c_{0} \epsilon, \\
    L_{5} &=& {\cal{N}} \frac{1}{4}\omega^{2}c_{0}^{2}, \\
    L_{6} &=& {\cal{N}} \left( - \frac{B^{2}}{2} \log \epsilon \right),
\end{eqnarray}
and we define the renormalized on-shell action as
\begin{equation}
    \tilde{S}_{D7}^{\rm ren} = \lim_{\epsilon \to 0}\left[ \tilde{S}_{D7} + \sum_{i=1}^{6} L_{i} \right].
\end{equation}
Note that $L_{3}$ and $L_{5}$ are finite in the limit of $\epsilon \to 0$.
In this work, we identify the thermodynamic potential with ${\rm \Omega}=-\tilde{S}_{D7}$, following the previous study \cite{OBannon:2008cmg}.
Here, we note that there is a subtlety in defining the thermodynamic potential for the steady state corresponding to a Minkowski embedding with a world-volume horizon, since no established thermodynamic framework for steady states currently exists. Nevertheless, we assume that the on-shell bulk action could holographically determine thermodynamic preference even for steady states, as it does in equilibrium. Previous studies have employed the renormalized on-shell action or the Legendre transformed action to evaluate thermodynamic stability of steady states \cite{Nakamura:2012ae,Imaizumi:2019byu,Matsumoto:2022nqu,Endo:2023uqk,Kundu:2013eba,Kundu:2019ull}.

\begin{figure}
    \centering
    \includegraphics[width=0.47\linewidth]{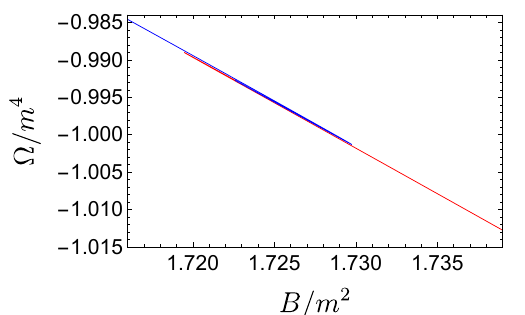}
    \includegraphics[width=0.47\linewidth]{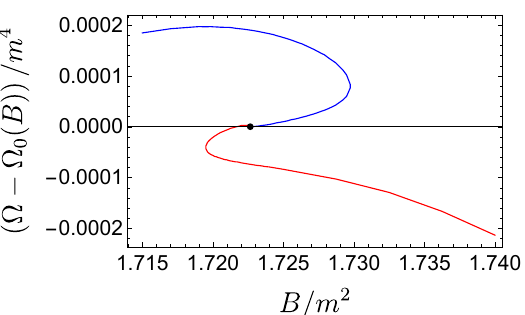}
    \caption{The thermodynamic potential ${\rm \Omega}/m^{4}$ as a function of $B/m^{2}$ for $\omega/m =1.00$. The Minkowski embedding with and without a world-volume horizon are denoted by blue and red colors, respectively. The right panel shows the deviations from the linear fit of the data ${\rm \Omega}_{0}(B)$ to make it easier to see.}
    \label{fig:Omega1}
\end{figure}

In the left panel of figure \ref{fig:Omega1}, we show the thermodynamic potential as a function of the magnetic field for $\omega/m=1.00$, corresponding to the plot in figure \ref{fig:BJplotT0} and \ref{fig:BcplotT0}. Since there are only slight differences around the critical embedding and the dependence appears almost linear, we also show the deviations from the linear fit of our numerical data, ${\rm \Omega}-{\rm \Omega_{0}}(B)$, in the right panel of figure \ref{fig:Omega1}. Note that the deviation at the critical embedding is set to zero by subtracting an appropriate constant. We find that the Minkowski embedding without a world-volume horizon is always thermodynamically preferred to that with a world-volume horizon.

\begin{figure}
    \centering
    \includegraphics[width=0.32\linewidth]{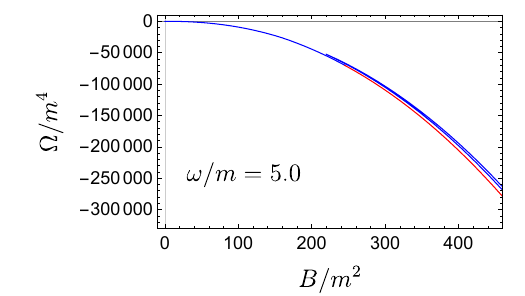}
    \includegraphics[width=0.32\linewidth]{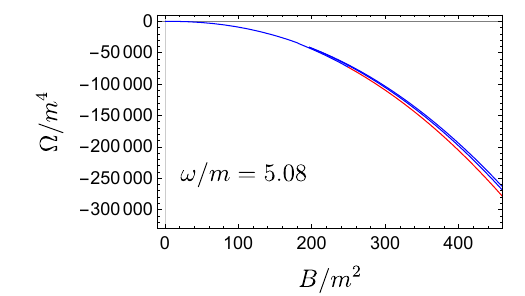}
    \includegraphics[width=0.32\linewidth]{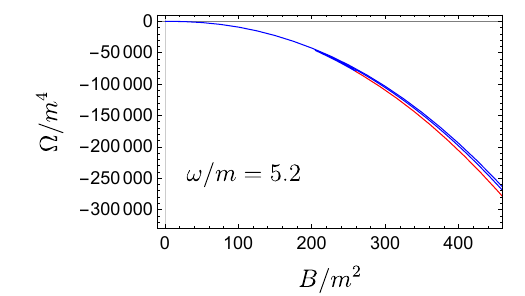}
    \includegraphics[width=0.32\linewidth]{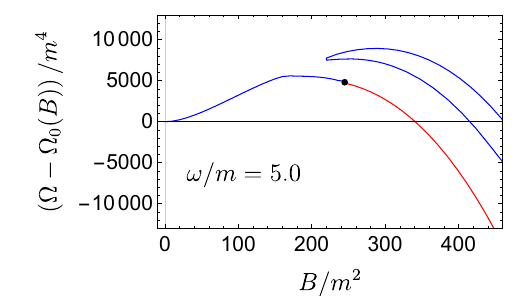}
    \includegraphics[width=0.32\linewidth]{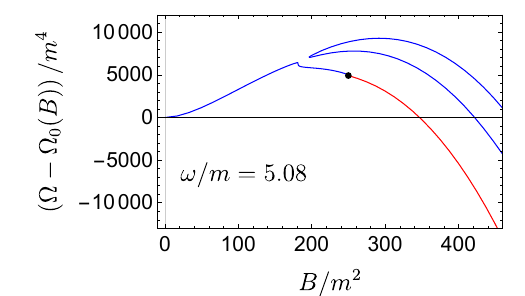}
    \includegraphics[width=0.32\linewidth]{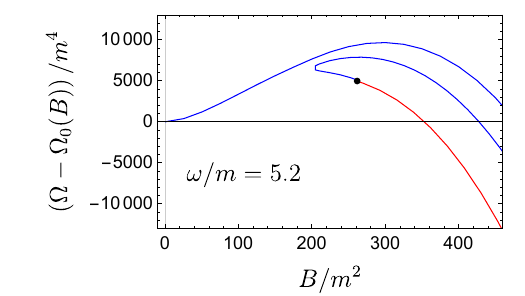}
    \caption{Upper panel: The thermodynamic potential ${\rm \Omega}/m^{4}$ as a function of $B/m^{2}$ for $\omega/m =5.0$ (left), $\omega/m=5.08$ (middle), and $\omega/m =5.2$ (right). The Minkowski embedding with and without a world-volume horizon are denoted by blue and red colors, respectively. Lower panel: The deviations from the quadratic fit of the data ${\rm \Omega}_{0}(B)$ to make it easier to see.}
    \label{fig:Omega2}
\end{figure}

We also compute the thermodynamic potential for large $\omega/m$, corresponding to figure \ref{fig:BJplotL}. In this case, the thermodynamic potential seems to be quadratic in $B/m^{2}$ as shown in the upper panel in figure \ref{fig:Omega2}. Then, we use a quadratic fit to our numerical data to show the deviations in the lower panel in figure \ref{fig:Omega2}.
In either case, we find that the Minkowski embedding without a world-volume horizon is always thermodynamically favored. At first glance, the plot of $\omega/m=5.2$ in figure \ref{fig:BJplotL} appears to suggest that the Minkwoski embedding with a world-volume horizon connects continuously as $B/m^{2}$ increases. The thermodynamic potential, however, shows that the branch with the smaller current density is thermodynamically favored. The transition between the Minkowski embedding with and without a world-volume horizon is always a first-order phase transition accompanied with the multi-valued behavior near the critical embedding. In contrast, if we regard the current density as an order parameter, we find that the first-order phase transition between different Minkowski embeddings with a world-volume horizon ($\omega/m = 5.2$) turns into a crossover ($\omega/m = 5.0$) as $\omega/m$ is varied. The intermediate case with $\omega/m=5.08$ may correspond to a critical endpoint, where the first-order transition terminates, since the derivative of the current density with respect to $B/m^{2}$ diverges at the critical value of $B/m^{2}$. This non-equilibrium phase transition between multiple steady states originates from the branch reconnection, and it would be interesting to compare this phenomenon with similar non-equilibrium phase transitions discussed in \cite{Nakamura:2012ae,Ali-Akbari:2013hba,Matsumoto:2018ukk,Endo:2023uqk}. 

%%%%%%%%%%%%%%
\section{Finite temperature} \label{sec:finiteT}
In this section, we study the chiral magnetic effect at finite temperatures. At finite temperature, as discussed in \cite{Hoyos:2011us}, another type of embedding solutions, referred to as black hole embeddings, are possible. For black hole embeddings, the D7-brane intersects the black hole horizon and the world-volume horizon necessarily emerges outside of the black hole horizon. 

\begin{figure}
    \centering
    \includegraphics[width=1.\linewidth]{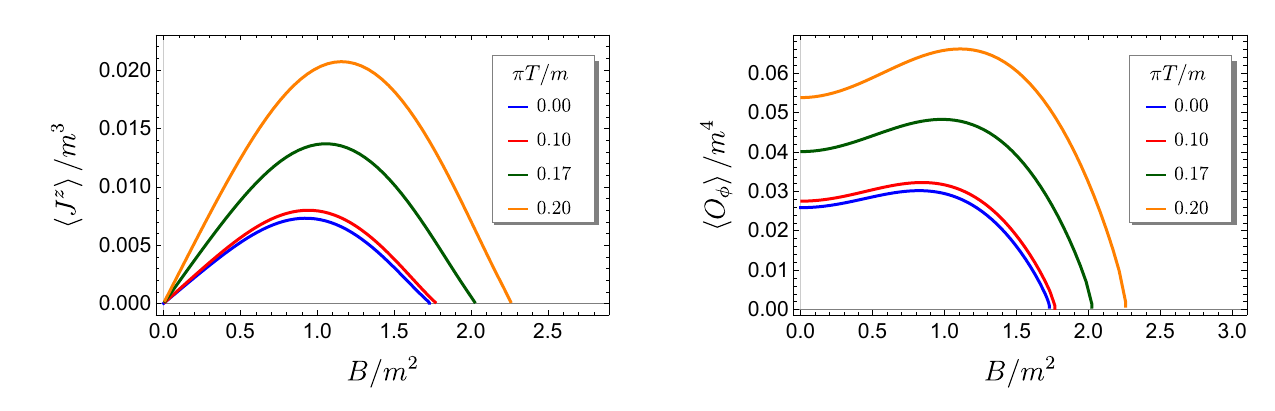}
    \caption{$\expval{J^{z}}/m^{3}$ and $\expval{\mathcal{O}_{\phi}}/m^{4}$ as a function of $B/m^{2}$ for several temperatures with $\omega/m =1.0$.}
    \label{fig:finiteTplot}
\end{figure}

In figure \ref{fig:finiteTplot}, we show $\expval{J^{z}}/m^{3}$ and $\expval{\mathcal{O}_{\phi}}/m^{4}$ as a function of $B/m^{2}$ for several temperatures with $\omega/m =1.0$, obtained from the black hole embeddings.
As temperature increases, the chiral magnetic current becomes larger for a given $B/m^{2}$, producing an effect qualitatively similar to increasing the angular frequency $\omega$ as in figure \ref{fig:BJplotT0}.
We do not observe additional qualitative changes or new phase structures at finite temperature within the explored parameter range, indicating that the qualitative features of the chiral magnetic response remain unchanged as temperature increases.
Our results are also consistent with those in \cite{Hoyos:2011us}.

%%%%%%%%%%%%%%
\section{Conclusions and discussions} \label{sec:conclusion}
In this work, we investigate the nonlinear response of the chiral magnetic current with respect to the external magnetic field in the D3/D7 model with a finite axial chemical potential. The holographic model has been originally proposed in \cite{Hoyos:2011us}, while we particularly focus on the critical behavior near the transition between the state with CME and without CME. Its transition originates from a series of solutions emerging in the vicinity of the conical singular solution in the bulk picture, discussed in this model with various setups \cite{Mateos:2006nu,Mateos:2007vn,Filev:2007gb,Albash:2006ew,Albash:2007bq,Erdmenger:2007bn,Frolov:2006tc,Ishigaki:2021vyv}. We find multiple solutions in the presence of the CME, namely solutions with different values of the chiral magnetic current for a given magnetic field exist (see figure \ref{fig:BJplotT0}). We confirm that they correspond to the same series of solutions around the conical singular solution, corresponding to the critical embedding in the bulk theory, by computing the chiral condensate for those solutions. We also characterize the near-critical behavior by extracting the scaling law and the discrete self-similar exponents.
Additionally, we discover a novel reconnection transition for two branches of solutions in the plot of the chiral magnetic current and magnetic field (see figure \ref{fig:BJplotL}).

From the viewpoint of dynamical stability for these solutions, we study the meson spectrum for the state without CME. As expected from the previous works, the solutions become dynamically unstable as they approach the critical embedding. However, the instability does not immediately appear at the turning point of the solution as observed in different setups \cite{Mateos:2006nu,Mateos:2007vn,Ishigaki:2021vyv}. As the angular frequency for the D7-brane increases, the onset of the instability moves toward the critical embedding (see figure \ref{fig:M2plot}). It implies that the presence of the axial chemical potential and the external magnetic field cooperatively stabilize the meson bound states. Although we focus on the meson spectrum in the Minkowski embedding without a world-volume horizon, investigating the Minkowski embedding with a world-volume horizon, as well as the black hole embedding, would be interesting. We leave this for future work.

Additionally, we discuss the thermodynamic stability determined by the thermodynamic potential, holographically corresponding to the renormalized on-shell action. We find that the Minkowski embedding without a world-volume horizon is thermodynamically favored compared to that with a world-volume horizon, and the transition between them is a first-order phase transition. For large $\omega/m$, where the branch reconnection occurs, the thermodynamically favored branch is found to be the one with the lower current density, suggesting that the first-order phase transition evolves into a crossover through a possible second-order critical point as $\omega/m$ is varied. It would be interesting to compare this non-equilibrium phase transition with similar phenomena discussed in \cite{Nakamura:2012ae,Ali-Akbari:2013hba,Matsumoto:2018ukk,Endo:2023uqk}.

Lastly, we investigate the chiral magnetic current and the pseudo-scalar condensate at finite temperature. Compared to the zero-temperature case, their dependence on the magnetic field remains qualitatively similar. By contrast, when a steady state is realized by applying an external electric field rather than an axial chemical potential, novel non-equilibrium phase transitions induced by temperature have been reported \cite{Nakamura:2012ae,Ali-Akbari:2013hba}. This distinction originates from the different mechanisms underlying the steady-state realization: a conducting state induced by an external electric field versus a CME state generated by an external magnetic field in the presence of an axial chemical potential.
From this perspective, switching on the external electric field along the same direction as the magnetic field could result in much richer phenomena, where the contribution of the anomaly given by $\vec{E}\cdot \vec{B}$ must be carefully considered \cite{Nakamura:2026fcr}.

Overall, our analysis uncovers a rich nonlinear structure of the chiral magnetic response in the holographic framework, characterized by multiple solution branches and their non-trivial stability properties.
The central result of this work is the emergence of a multivalued nonlinear response associated with the critical embedding, leading to multiple steady states and non-equilibrium phase transitions. These findings provide a new perspective on the role of critical embeddings in holographic chiral transport and on the thermodynamic and dynamical properties of non-equilibrium steady states.

\section*{Acknowledgements}
We thank the anonymous referee for constructive comments and valuable suggestions that significantly improved the manuscript.
The work of MM is supported by Shanghai Sci-tech Co-research Program (Grant No. 25HB2700600).

\appendix
\section{Equations for fluctuations} \label{sec:AppA}
In this appendix, we show the explicit forms of the equations for fluctuations $\{\delta \tilde{R},\delta \tilde{\varphi}, \delta \tilde{A_{z}} \}$.
For brevity we omit tildes in the following. We write the equations as the form of
\begin{eqnarray}
    \delta R'' + a_{R,1} \delta R' + a_{R,2} \delta \varphi' + a_{R,3} \delta A_{z}' + b_{R,1} \delta R+ b_{R,2} \delta \varphi + b_{R,3} \delta A_{z} &=& 0, \\
    \delta \varphi'' + a_{\varphi,1} \delta R' + a_{\varphi,2} \delta \varphi' + a_{\varphi,3} \delta A_{z}' + b_{\varphi,1} \delta R+ b_{\varphi,2} \delta \varphi + b_{\varphi,3} \delta A_{z} &=& 0, \\
    \delta A_{z}'' + a_{A,1} \delta R' + a_{A,2} \delta \varphi' + a_{A,3} \delta A_{z}' + b_{A,1} \delta R+ b_{A,2} \delta \varphi + b_{A,3} \delta A_{z} &=& 0, 
\end{eqnarray}
and define
\begin{eqnarray}
    P &=& 1 + u^{4} R',\\
    L &=& \left(B^2 u^4+R^4 u^4+2 R^2 u^2+1\right) \left(R^4 u^4+R^2 \left(2 u^2-u^4 \omega ^2\right)+1\right).
\end{eqnarray}
Then, the coefficients of the equation for $\delta R$ are
\begin{align}
    a_{R,1} ={} & \Big[
      P R^5 R' u^9 \left(\omega^2 - 2 B^2 u^2\right) \notag\\
    & + P R^3 R' u^7 \left(B^2 u^2 (3 u^2 \omega^2 - 4) - \omega^2\right) \notag\\
    & - P R R' u^5 \left(B^2 u^2 (u^2 \omega^2 + 2) + \omega^2\right) \notag\\
    & - R^6 u^6 \Big(B^2 \left(7 R'^2 u^8 + 3 R' R'' u^9 + u^4\right)
        + R'^2 \left(70 u^4 - 19 u^6 \omega^2\right)
        + R' R'' \left(30 u^5 - 9 u^7 \omega^2\right) - u^2 \omega^2 + 10\Big) \notag\\
    & + R^4 u^4 \Big(B^2 u^4 \left(R'^2 u^4 (7 u^2 \omega^2 - 19)
        + 3 R' R'' u^5 (u^2 \omega^2 - 3) + u^2 \omega^2 - 1\right) \notag\\
    & \hspace{2.7em} + R'^2 u^4 (17 u^2 \omega^2 - 70)
        + 3 R' R'' u^5 (3 u^2 \omega^2 - 10) - u^2 \omega^2 - 10\Big) \notag\\
    & + R^2 u^2 \Big(B^2 \left(R'^2 u^8 (3 u^2 \omega^2 - 17)
        + 3 R' R'' u^9 (u^2 \omega^2 - 3) - 3 u^6 \omega^2 + u^4\right) \notag\\
    & \hspace{2.7em} + 5 R'^2 u^4 (u^2 \omega^2 - 7)
        + 3 R' R'' u^5 (u^2 \omega^2 - 5) - u^2 \omega^2 - 5\Big) \notag\\
    & + B^2 \left(-5 R'^2 u^8 - 3 R' R'' u^9 + u^4\right) \notag\\
    & + P R^7 R' u^{11} \omega^2 \notag\\
    & - R^{10} u^{10} \left(7 R'^2 u^4 + 3 R' R'' u^5 + 1\right) \notag\\
    & + R^8 u^8 \left(u^2 \omega^2 - 5\right) \left(7 R'^2 u^4 + 3 R' R'' u^5 + 1\right) \notag\\
    & - 7 R'^2 u^4 - 3 R' R'' u^5 - 1
    \Big] \Big/ \left(u \left(1 + R^2 u^2\right) L P\right), \label{eq:aR1}\\[1em]
    a_{R,2} ={}& 0, \\[0.3em]
    a_{R,3} ={}& \frac{4 B P^2 R u^3 \omega}{\left(1 + R^2 u^2\right)\sqrt{L P}},
\end{align}
and
\begin{align}
    b_{R,1} ={} & \bigg[
    B^4 \Big(R \Big(R'^3 \big(6 R^{10} u^{10} + R^8 (22 - 15 u^2 \omega^2) u^8 + R^6 (9 u^4 \omega^4 - 12 u^2 \omega^2 + 28) u^6 \notag\\
    & \hspace{3em} - 2 R^4 (7 u^4 \omega^4 - 9 u^2 \omega^2 - 6) u^4 - 3 \omega^2 u^2 + R^2 (u^4 \omega^4 + 12 u^2 \omega^2 - 2) u^2 - 2\big) u^4 \notag\\
    & \hspace{2em} - R'' (R^2 u^2 + 1) \big(2 R^8 u^8 + 3 R^4 (u^4 \omega^4 - 3 u^2 \omega^2 + 4) u^4 + \omega^2 u^2 \notag\\
    & \hspace{3em} - R^2 (u^4 \omega^4 + 3 u^2 \omega^2 - 8) u^2 + R^6 (8 u^6 - 5 u^8 \omega^2) + 2\big) u \notag\\
    & \hspace{2em} + R' \big(2 R^{10} u^{10} + R^8 (2 - 5 u^2 \omega^2) u^8 + R^6 (3 u^4 \omega^4 + 16 u^2 \omega^2 - 12) u^6 \notag\\
    & \hspace{3em} - 2 R^4 (9 u^4 \omega^4 - 21 u^2 \omega^2 + 14) u^4 - 5 \omega^2 u^2 + R^2 (3 u^4 \omega^4 + 16 u^2 \omega^2 - 22) u^2 - 6\big)\Big) u^3 \notag\\
    & \hspace{1em} + P \big(-6 R^{10} u^{10} + R^8 (19 u^2 \omega^2 - 22) u^8 - 2 R^6 (6 u^4 \omega^4 - 11 u^2 \omega^2 + 14) u^6 \notag\\
    & \hspace{3em} + 12 R^4 (u^4 \omega^4 - u^2 \omega^2 - 1) u^4 + \omega^2 u^2 + R^2 (2 u^2 - 14 u^4 \omega^2) + 2\big)\Big) u^6 \notag\\
    & + k_0^2 P (R^2 u^2 + 1)^2 (R^4 u^4 + B^2 u^4 + 2 R^2 u^2 + 1)^2 (R^4 u^4 + R^2 (2 u^2 - u^4 \omega^2) + 1) \notag\\
    & + (R^2 u^2 + 1)^4 \omega^2 \Big(R \Big(-R'^3 \big(3 R^8 u^8 - R^6 (3 u^2 \omega^2 + 2) u^6 + 6 R^4 (u^2 \omega^2 - 2) u^4 \notag\\
    & \hspace{3em} + R^2 (u^2 \omega^2 - 6) u^2 + 1\big) u^4 + R'' (R^4 u^4 - 1)(R^4 u^4 + R^2 (2 u^2 - u^4 \omega^2) + 1) u \notag\\
    & \hspace{2em} + R' \big(-R^8 u^8 + R^6 (u^2 \omega^2 + 6) u^6 - 6 R^4 (u^2 \omega^2 - 2) u^4 + R^2 (u^2 \omega^2 + 2) u^2 - 3\big)\Big) u^3 \notag\\
    & \hspace{1em} + P \big(3 R^8 u^8 + 6 R^4 (u^2 \omega^2 - 2) u^4 - 6 R^2 u^2 - 2 R^6 (\omega^2 u^8 + u^6) + 1\big)\Big) \notag\\
    & - 2 B^2 (R^2 u^3 + u)^2 \Big(R \Big(R'^3 \big(-3 R^{10} u^{10} + 9 R^8 (u^2 \omega^2 - 1) u^8 + R^6 (-6 u^4 \omega^4 + u^2 \omega^2 - 6) u^6 \notag\\
    & \hspace{3em} + R^4 (12 u^4 \omega^4 - 23 u^2 \omega^2 + 6) u^4 + 2 \omega^2 u^2 + R^2 (9 u^2 - 13 u^4 \omega^2) + 3\big) u^4 \notag\\
    & \hspace{2em} + R'' (R^2 u^2 + 1) \big(R^8 u^8 + R^4 (2 u^4 \omega^4 - 5 u^2 \omega^2 + 6) u^4 + \omega^2 u^2 \notag\\
    & \hspace{3em} - R^2 (u^4 \omega^4 + u^2 \omega^2 - 4) u^2 + R^6 (4 u^6 - 3 u^8 \omega^2) + 1\big) u \notag\\
    & \hspace{2em} + R' \big(-R^{10} u^{10} + R^6 (-2 u^4 \omega^4 - 15 u^2 \omega^2 + 14) u^6 + R^4 (14 u^4 \omega^4 - 35 u^2 \omega^2 + 26) u^4 \notag\\
    & \hspace{3em} + 4 \omega^2 u^2 + R^8 (3 \omega^2 u^{10} + u^8) + R^2 (-2 \omega^4 u^6 - 13 \omega^2 u^4 + 19 u^2) + 5\big)\Big) u^3 \notag\\
    & \hspace{1em} + P \big(5 R^{10} u^{10} + R^8 (19 - 15 u^2 \omega^2) u^8 + R^6 (9 u^4 \omega^4 - 18 u^2 \omega^2 + 26) u^6 - \omega^2 u^2 \notag\\
    & \hspace{3em} + R^2 (10 \omega^2 u^4 + u^2) + R^4 (-9 \omega^4 u^8 + 8 \omega^2 u^6 + 14 u^4) - 1\big)\Big)
    \bigg]
    \bigg/ \Big[(1+R^2 u^2)^2 L^2\Big], \label{eq:bR1} \\[1em]
    b_{R,2} ={} & - \bigg[
    i k_0 R \omega \Big(P \big(R^6 u^6 (-4 B^2 u^4 + u^2 \omega^2 - 4) + R^4 u^4 (3 B^2 u^4 (u^2 \omega^2 - 2) - u^2 \omega^2 + 4) \notag\\
    & \hspace{2em} - R^2 u^2 (B^2 u^6 \omega^2 + u^2 \omega^2 - 6) + 2 B^2 u^4 + R^8 u^8 (u^2 \omega^2 - 6) + 2\big) \notag\\
    & \hspace{1em} + R u^3 \Big(-u \big(R'' (R^2 u^2 + 1)(B^2 u^4 + R^4 u^4 + 2 R^2 u^2 + 1)(R^4 u^4 + R^2 (2 u^2 - u^4 \omega^2) + 1) + 2 R^9 u^6\big) \notag\\
    & \hspace{2em} + R'^3 u^4 \big(R^6 u^6 (3 B^2 u^4 - 7 u^2 \omega^2 + 14) + R^4 u^4 (B^2 (3 u^4 - 3 u^6 \omega^2) - 5 u^2 \omega^2 + 6) \notag\\
    & \hspace{3em} + R^2 u^2 (B^2 u^4 (u^2 \omega^2 - 3) - u^2 \omega^2 - 1) - 3 B^2 u^4 + 3 R^{10} u^{10} + R^8 u^8 (11 - 3 u^2 \omega^2) - 1\big) \notag\\
    & \hspace{2em} + R' \big(R^6 u^6 (B^2 u^4 - u^2 \omega^2 - 6) - R^4 u^4 (B^2 u^4 (u^2 \omega^2 + 3) - u^2 \omega^2 + 14) \notag\\
    & \hspace{3em} + R^2 u^2 (3 B^2 u^4 (u^2 \omega^2 - 3) + u^2 \omega^2 - 11) - 5 B^2 u^4 + R^{10} u^{10} + R^8 (u^8 - u^{10} \omega^2) - 3\big) \notag\\
    & \hspace{2em} - 2 R^9 R'^2 u^{11}\Big)\Big)
    \bigg]
    \bigg/ \Big[(1+R^2 u^2 )(R^4 u^4 + R^2 (2 u^2 - u^4 \omega^2) + 1)L\Big], \\[0.3em]
    b_{R,3} ={} & 0.
\end{align}
The coefficients of equation for $\delta \varphi$ are
\begin{align}
    a_{\varphi,1} ={} & 0, \\
    a_{\varphi,2} ={} & - \bigg[
    P R^2 R' u^3 \left(B^2 u^4 (u^2 \omega^2 - 2) + u^2 \omega^2 - 8\right)
    - 2 P R' u (B^2 u^4 + 1) \notag\\
    & + R^5 u^4 \Big(R'^2 u^4 (3 B^2 u^4 - 8 u^2 \omega^2 + 18) + R' R'' u^5 (B^2 u^4 - 2 u^2 \omega^2 + 6) + B^2 u^4 - 4 u^2 \omega^2 + 6\Big) \notag\\
    & - R^3 u^2 \Big(R'^2 u^4 \left(B^2 u^4 (3 u^2 \omega^2 - 4) + 5 u^2 \omega^2 - 12\right) + R' R'' u^5 \left(B^2 u^4 (u^2 \omega^2 - 2) + u^2 \omega^2 - 4\right) \notag\\
    & \hspace{2em} + B^2 u^6 \omega^2 + 3 u^2 \omega^2 - 4\Big) \notag\\
    & + R \left(B^2 u^4 (R'^2 u^4 + R' R'' u^5 - 1) + 3 R'^2 u^4 + R' R'' u^5 + 1\right) \notag\\
    & - 2 P R^8 R' u^9 + P R^6 R' u^7 (3 u^2 \omega^2 - 8) + 4 P R^4 R' u^5 (u^2 \omega^2 - 3) \notag\\
    & + R^9 u^8 \left(3 R'^2 u^4 + R' R'' u^5 + 1\right) - R^7 u^6 (u^2 \omega^2 - 4)\left(3 R'^2 u^4 + R' R'' u^5 + 1\right)
    \bigg] \bigg/ \big[P R u L\big], \\
    a_{\varphi,3} ={} & 0,
\end{align}
and
\begin{align}
    b_{\varphi,1} ={} & \bigg[
    i k_0 \omega \Big(P \big(R^6 u^6 (-4 B^2 u^4 + u^2 \omega^2 - 4) + R^4 u^4 (3 B^2 u^4 (u^2 \omega^2 - 2) - u^2 \omega^2 + 4) \notag\\
    & \hspace{2em} - R^2 u^2 (B^2 u^6 \omega^2 + u^2 \omega^2 - 6) + 2 B^2 u^4 + R^8 u^8 (u^2 \omega^2 - 6) + 2\big) \notag\\
    & \hspace{1em} + R u^3 \Big(-u \big(R'' (R^2 u^2 + 1)(B^2 u^4 + R^4 u^4 + 2 R^2 u^2 + 1)(R^4 u^4 + R^2 (2 u^2 - u^4 \omega^2) + 1) + 2 R^9 u^6\big) \notag\\
    & \hspace{2em} + R'^3 u^4 \big(R^6 u^6 (3 B^2 u^4 - 7 u^2 \omega^2 + 14) + R^4 u^4 (B^2 (3 u^4 - 3 u^6 \omega^2) - 5 u^2 \omega^2 + 6) \notag\\
    & \hspace{3em} + R^2 u^2 (B^2 u^4 (u^2 \omega^2 - 3) - u^2 \omega^2 - 1) - 3 B^2 u^4 + 3 R^{10} u^{10} + R^8 u^8 (11 - 3 u^2 \omega^2) - 1\big) \notag\\
    & \hspace{2em} + R' \big(R^6 u^6 (B^2 u^4 - u^2 \omega^2 - 6) - R^4 u^4 (B^2 u^4 (u^2 \omega^2 + 3) - u^2 \omega^2 + 14) \notag\\
    & \hspace{3em} + R^2 u^2 (3 B^2 u^4 (u^2 \omega^2 - 3) + u^2 \omega^2 - 11) - 5 B^2 u^4 + R^{10} u^{10} + R^8 (u^8 - u^{10} \omega^2) - 3\big) \notag\\
    & \hspace{2em} - 2 R^9 R'^2 u^{11}\Big)\Big)
    \bigg] \bigg/ \Big[P R (R^2 u^2 + 1)^3 (B^2 u^4 + R^4 u^4 + 2 R^2 u^2 + 1)(R^4 u^4 + R^2 (2 u^2 - u^4 \omega^2) + 1)\Big], \\[1em]
    b_{\varphi,2} ={} & \frac{k_0^2 P}{R^4 u^4 + R^2 (2 u^2 - u^4 \omega^2) + 1}, \\[0.3em]
    b_{\varphi,3} ={} & - \frac{4 i B k_0 u^2 (R + R' u) }{R (R^2 u^2 + 1)^3}\sqrt{\dfrac{P \left(R^4 u^4 + R^2 (2 u^2 - u^4 \omega^2) + 1\right)}{B^2 u^4 + R^4 u^4 + 2 R^2 u^2 + 1}}.
\end{align}
The coefficients of equation for $\delta A_{z}$ are
\begin{align}
    a_{A,1} ={} & - \frac{4 B P R u^3 \omega}{(1+R^2 u^2 ) \sqrt{P L}}, \\[0.3em]
    a_{A,2} ={} & 0, \\[0.3em]
    a_{A,3} ={} & \bigg[
    P R^5 R' u^9 (\omega^2 - 2 B^2 u^2) + P R^3 R' u^7 \left(B^2 u^2 (3 u^2 \omega^2 - 4) - \omega^2\right) \notag\\
    & - P R R' u^5 \left(B^2 u^2 (u^2 \omega^2 + 2) + \omega^2\right) \notag\\
    & - R^6 u^6 \Big(B^2 (3 R'^2 u^8 + R' R'' u^9 + u^4) + R'^2 (30 u^4 - 7 u^6 \omega^2) + R' R'' u^5 (10 - 3 u^2 \omega^2) - u^2 \omega^2 + 10\Big) \notag\\
    & + R^4 u^4 \Big(B^2 u^4 \left(R'^2 u^4 (3 u^2 \omega^2 - 7) + R' R'' u^5 (u^2 \omega^2 - 3) + u^2 \omega^2 - 1\right) \notag\\
    & \hspace{2em} + 5 R'^2 u^4 (u^2 \omega^2 - 6) + R' R'' u^5 (3 u^2 \omega^2 - 10) - u^2 \omega^2 - 10\Big) \notag\\
    & + R^2 u^2 \Big(B^2 \left(-R'^2 u^8 (u^2 \omega^2 + 5) + R' R'' u^9 (u^2 \omega^2 - 3) - 3 u^6 \omega^2 + u^4\right) \notag\\
    & \hspace{2em} + R'^2 u^4 (u^2 \omega^2 - 15) + R' R'' u^5 (u^2 \omega^2 - 5) - u^2 \omega^2 - 5\Big) \notag\\
    & + B^2 \left(-R'^2 u^8 - R' R'' u^9 + u^4\right) + P R^7 R' u^{11} \omega^2 \notag\\
    & - R^{10} u^{10} \left(3 R'^2 u^4 + R' R'' u^5 + 1\right) + R^8 u^8 (u^2 \omega^2 - 5)\left(3 R'^2 u^4 + R' R'' u^5 + 1\right) \notag\\
    & - 3 R'^2 u^4 - R' R'' u^5 - 1
    \bigg] \bigg/ \Big[P u (1+R^2 u^2 )L\Big],
\end{align}
and
\begin{align}
    b_{A,1} ={} &  \frac{4 B P u^2 \omega  \left(4 R^3 u^2+5 R^2 R' u^3-2 R-R' u\right)}{\left(R^2 u^2+1\right)^2 \sqrt{P L}}\\[0.3em]
    b_{A,2} ={} & \frac{4 i B k_{0} P R u^2 (R+R' u)}{\left(R^2 u^2+1\right) \sqrt{P L}}, \\[0.3em]
    b_{A,3} ={} & \frac{k_{0}^2 P}{R^4 u^4+R^2 \left(2 u^2-u^4 \omega ^2\right)+1}.
\end{align}

\bibliography{main}
\bibliographystyle{JHEP}

\end{document}